\documentclass[10pt,leqno,oneside]{article}
\usepackage{amsmath, amssymb}
\usepackage{amsfonts}
\newcommand{\field}[1]{\mathbb{#1}}
\newcommand{\bn}{\mbox{\boldmath$\nabla$}}

\newcommand{\n}{\noindent}
\newcommand{\be}{\begin{equation}}
\newcommand{\ee}{\end{equation}}
\newcommand{\ben}{\begin{displaymath}}
\newcommand{\Su}{{\mathcal{S}}}
\newcommand{\een}{\end{displaymath}}
\newcommand{\ep}{\hspace{\stretch{1}}$\Box$}

\newcommand{\vs}{\vspace{0.2cm}}
\newcommand{\dt}{{\bf \Pi}}

\usepackage{epsfig}
\usepackage{latexsym}
\usepackage{mathrsfs}

\newtheorem{Proposition}{Proposition}

\newtheorem{Conjecture}{Conjecture}
\newtheorem{Lemma}{Lemma}

\newtheorem{Corollary}{Corollary}
\newtheorem{Example}{Example}
\newtheorem{Problem}{Problem}

\begin{document}

\begin{center}
{\Large\bf Scalar curvature, isoperimetric collapse and General Relativity in the Constant Mean Curvature gauge.}

\vspace{0.4cm}

{\large Martin Reiris}\footnote{e-mail: martin@aei.mpr.de}\\

\vspace{0.1cm}

\textsc{Albert Einstein Institute - Max Planck.}\\

\end{center}

\vspace{0.3cm}
\begin{center}
\begin{minipage}[c]{11cm}
\linespread{.9}%
\selectfont
{\small 
We discuss a set of relations, set in the form of results, conjectures and problems, between the $L^{2}_{g}$-norm of the Ricci curvature of a 3-manifold, the scalar curvature and the volume radius. We illustrate the scope of these relations with potential applications to the Einstein Constant Mean Curvature flow (or GR seen as a geometric flow of constant mean curvature), but we believe the framework has it own geometric interest.}

\vspace{1cm}
{\center \tableofcontents}

\end{minipage}
\end{center}

\newpage
{\center \section{Introduction.}\label{S1}}
{\it In this article we will work exclusively in dimension three. The results and conjectures do not have natural extensions to higher dimensions.}  

In modern approaches to Riemannian geometry it is standard to use intrinsic diffemorphism invariant quantities to control or understand the geometry (see for instance \cite{Petersen}). Examples of such invariant quantities are for instance, the volume, the diameter, the signature of the sectional curvature, lower bounds or particular norms of it. A quite used settup is the well established theory of Riemannian spaces with sectional curvature uniformly bounded above, i.e. when there is an a priori upper bound on $\|Ric\|_{L^{\infty}_{g}}$ (for related topics, definitions and references see\cite{Petersen}). Two basic statements in this theory are: 

\vs
{\it (Convergence I). A pointed sequence (of complete spaces) $\{(\Sigma_{i},g_{i},o_{i})\}$ with $\|Ric\|_{L^{\infty}_{g}(\Sigma_{i})}\leq \Lambda_{0}$, $inj_{g_{i}}(\Sigma_{i})\geq i_{0}>0$, has a subsequence converging in $C^{1,\beta}$ to a complete space $(\tilde{\Sigma},\tilde{g},\tilde{o})$.} 

\vs
\n This statement can be relaxed to 

\vs
{\it (Convergence II). A pointed sequence (of complete spaces) $\{(\Sigma_{i},g_{i},o_{i})\}$ with $\|Ric\|_{L^{\infty}_{g}(\Sigma_{i})}\leq \Lambda_{0}$, $inj_{g_{i}}(o_{i})\geq i_{0}>0$, has a subsequence converging in $C^{1,\beta}$ to a complete space $(\tilde{\Sigma},\tilde{g},\tilde{o})$.} 

\vs
Some remarks on these statements are in order. First, if instead of an a priori $L^{\infty}_{g}$-bound in the {\it Convergence} cases {\it I} and {\it II}, one has a priori $C^{i,\alpha}_{g}$-bounds on $Ricci$, then the metric $\tilde{g}$ is known to have $C^{i,\beta}$ regularity, $\beta<\alpha$ (the convergence is in $C^{i,\beta}$). Secondly an essential ingredient to show in the {\it Convergence} case {\it II} that the metric $\tilde{g}$ is complete is the fact that the reciprocal of the injectivity radius at a point $p$, $1/inj(p)$, is controlled from above by and upper bound on $\|Ric\|_{L^{\infty}_{g}}$, $1/inj(o)$, and $dist_{g}(p,o)$. This property can be seen indeed as a direct consequence of the Bishop-Gromov volume comparison (under lower bounds on the Ricci curvature)
\footnote{Suppose the sectional curvature is bounded below by $K$. The Bishop-Gromov volume comparison implies that the volume quotient $Vol_{g}(B_{g}(p,s))/Vol_{g_{K}}(B(\bar{o},s))$ where $Vol_{g_{K}}(\bar{o},s))$ is the volume of a ball of radius $s$ and origin $\bar{o}$ in the space form of sectional curvature $K$, is monotonically decreasing as $s$ increases. Therefore, as there is some volume around $o$, we have that the quotient for $s=d(p,o)$ is less than the quotient for all $s\leq \inf\{dist_{g}(p,o),1\}$ thus we have $Vol_{g}(B(p,s))\geq c(dist_{g}(p,o),inj(o),K,\|Ric\|_{L^{\infty}_{g}}) s^{3}$. Thus the volume radius is controlled from below at a finite distance from the non-collapsed point $o$. Finally an upper bound on $\|Ric\|_{L^{\infty}_{g}}$ and a lower bound on the volume radius at a point gives a lower bound on the injectivity radius at the point (proceed by contradiction).}. We can phrase it by saying that, under a priori $L^{\infty}_{g}$ curvature bounds, there cannot be {\it collapse at a finite distance form a non-collapsed region}, where by this we mean that $inj(p)$ cannot get arbitrarily small if $\|Ric\|_{L^{\infty}_{g}}\leq \Lambda_{0}$, $dist_{g}(p,o)\leq d_{0}$ and $inj(o)\geq i_{0}>0$.  

Although the discussion of the $L^{s}_{g}$-theory, when $s>3/2$, (i.e. when there is an a priori upper bound on the $L^{s}_{g}$-norm of the Ricci curvature) goes essentially in parallel to the $L^{\infty}_{g}$($C^{\alpha}_{g}$)-theory, there are differences and some of them substantial. Let us explain this in more detail. In the $L^{s}_{g}$-theory the right notion in substitution of the injectivity radius is the {\it volume radius}. The (well known) definition of the volume radius proceeds like this. Let $(\Sigma,g)$ be a Riemannian-three manifold. We do not necessarily assume that $(\Sigma,g)$ is complete, or that it is with or without boundary. Given $\delta>1$ (fixed) define the {\it $\delta$-volume radius} at a point $\{p\}$ (to be denoted by $\nu^{\delta}(p)$), as
\ben
\nu^{\delta}(p)=\sup\{r<Rad(p)/ \forall B(q,s)\subset B(p,r),\ \frac{1}{\delta^{3}}w_{1}s^{3}\leq Vol(B(p,s))\leq \delta^{3}w_{1}s^{3}\},
\een

\n where $Rad(p)$ is the radius at $p$ and is defined as the distance from $p$ to the boundary of the metric completion of $(\Sigma,g)$. The boundary of $\Sigma$ is the metric completion minus the set $\Sigma$. $w_{1}$ is the volume of the unit three-sphere. The {\it volume radius of a region $\Omega$} denoted by $\nu^{\delta}(\Omega)$ is defined as the infimum of $\nu^{\delta}(p)$ when $p$ varies in $\Omega$. The main statement linking the $L^{s}_{g}$-norm of Ricci and the volume radius is the following: {\it Let $(\Sigma,g)$ be a complete Riemannian three-manifold, let $r>0$ and let $p$ be an arbitrary point in $\Sigma$. Then  the reciprocal of the $H^{2}$-harmonic radius $r_{H}^{2}(p)$ at $p$ is controlled from above by an upper bound on $1/r$, $\|Ric\|_{L^{s}_{g}(B_{g}(p,r))}$ and $\nu^{\delta}(p)$.} The harmonic radius is defined as the supremum of the radiuses $r>0$ for which there is a harmonic coordinate chart $\{x\}$ covering $B_{g}(p,r)$ and satisfying
\begin{equation}\label{HCHI}
\frac{3}{4}\delta_{jk}\leq g_{jk}\leq \frac{4}{3}\delta_{jk},
\end{equation}
\begin{equation}\label{HCHII}
r(\sum_{|I|=2,j,k}\int_{B(p,r)}|\frac{\partial^{I}}{\partial x^{I}}g_{jk}|^{2}dv_{x})\leq 1,
\end{equation}

\noindent where in the sum above $I$ is the multindex $I=(\alpha_{1},\alpha_{2},\alpha_{3})$, and as usual 

\n $\partial^{I}/\partial x^{I}=(\partial_{x^{1}})^{\alpha_{1}}(\partial_{x^{2}})^{\alpha_{2}}(\partial_{x^{3}})^{\alpha_{3}}$. Note that both expressions above are invariant under the simultaneous scaling $\tilde{g}=\lambda^{2}g$, $\tilde{x}^{\mu}=\lambda x^{\mu}$ and $\tilde{r}=\lambda r$. Thus, in a sense, the harmonic $H^{2}$-radius controls the $H^{2,s}_{\{x\}}$-Sobolev norm of $g_{ij}-\delta_{ij}$ in a suitable harmonic chart $\{x\}$. This fact is the root of a basic {\it Convergence I} statement in the $L^{s}_{g}$-theory 

\vs
{\it (Convergence I)}. A pointed sequence $\{(\Sigma_{i},g_{i},o_{i})\}$ of complete Riemannian three-manifolds with $\|Ric\|_{L^{s}_{g_{i}}(\Sigma_{i})}\leq \Lambda_{0}$, ($s>3/2$), and $\nu^{\delta}(\Sigma_{i})\geq \nu_{0}>0$  has a subsequence converging in the weak $H^{2}$-topology to a complete Riemannian manifold $(\tilde{\Sigma},\tilde{g},\tilde{o})$.

\vs 
In spite of this, there is one significant difference between the $L^{s}_{g}$ and the $L^{\infty}_{g}$($C^{\alpha}_{g}$)-theories which is largely due to the lack of volume comparison in this context: the property of {\it non-collpase at a finite distance from a non-collapsed region} does not hold in general. As a consequence the statement {\it Convergence II} which is true in the $L^{\infty}_{g}$($C^{\alpha}_{g}$)-theory is false in the $L^{s}_{g}$-theory.  There is instead a relative but weaker property in the $L^{s}_{g}$-theory saying that {\it there is non-collapse in distances less than a definite distance from a non-collapsed region}. This is the content of the following result of D. Yang \cite{Yang}:  {\it There is $d_{0}(\Lambda_{0},\nu_{0})>0$ such that for any complete pointed space $(\Sigma,g, o)$ with $\|Ric\|_{L^{s}_{g}(\Sigma)}\leq \Lambda_{0}$, $s>3/2$, and $\nu^{\delta}(o)\geq \nu_{0}>0$ then $\nu^{\delta}(p)\geq \nu(\Lambda_{0},\nu_{0})>0$ provided $dist_{g}(p,o)\leq d_{0}$. Moreover for any sequence of spaces where $\nu_{0}$ remains fixed but $\Lambda_{0}\rightarrow 0$, we have $d_{0}\rightarrow \infty$.} The next two classical examples (also due to D. Yang) show that the phenomena of collapse at finite distances can indeed occur. 

\vs
\begin{Example}\label{EXC1}{\rm Let $g_{\epsilon}=dx^{2}+(\epsilon+x^{2})^{4}(d\theta_{1}^{2}+d\theta_{2}^{2})$ be a family of metrics on $[-1,1]\times S^{1}\times S^{1}$. When $\epsilon\rightarrow 0$ the volume radius at the central torus $\{x=0\}$ collapses to zero, while the volume radius at any point in the torus $\{x=1\}$ remains uniformly bounded below. The integral curvature $\|Ric\|_{L^{2}_{g}}$ remains uniformly bounded above but the scalar curvature diverges to minus infinity at the central torus.}
\end{Example}

\begin{Example}\label{EXC2}{\rm Let $g_{\epsilon}=dr^{2}+r^{2} d\theta_{2}^{2}+(\epsilon+r^{2})^{4}d\theta_{1}^{2}$ be a family of metrics in the manifold $D^{2}\times S^{1}$ where $D^{2}$ is the two-dimensional unit disc. As $\epsilon\rightarrow 0$ the volume radius at the central circle $\{r=0\}$ collapses to zero, while the volume radius at any point in the torus $\{r=1\}$ remains uniformly bounded below. The integral curvature $\|Ric\|_{L^{2}_{g}}$ remains uniformly bounded above but the scalar curvature at the central circle diverges to negative infinity.} 
\end{Example}

\n An important consequence of the result of D. Yang that we have stated before is a theorem of pointed convergence but into a particular family of (potentially) incomplete Riemannian spaces. To make this precise let us introduce some definitions. Let $(\Sigma,g)$ be a non-complete Riemannian manifold, {\it we say that $(\Sigma,g)$ is $\nu^{\delta}$-complete\footnote{This terminology is ours.} iff $\nu^{\delta}(\Sigma)=0$ and $Rad(p_{i})/\nu^{\delta}(p_{i})\rightarrow \infty$ when $Rad(p_{i})\rightarrow 0$}. The intuition is that these type of manifolds are complete at the scale of $\nu^{\delta}$, namely that when one scales the metric $g$ by $1/\nu^{\delta}(p_{i})^{2}$ then the boundary of $\Sigma_{i}$ lies further and further away from $p_{i}$ as $\nu^{\delta}(p_{i})\rightarrow 0$. Note that in the metric $\tilde{g}_{p_{i}}=\frac{1}{\nu^{\delta}(p_{i})^{2}}g$ it is $\nu^{\delta}_{\tilde{g}_{p_{i}}}(p_{i})=1$. 

\vs
{\it (Convergence II - $L^{s}_{g}$-theory). Let $\{(\Sigma_{i},g_{i},p_{i})\}$ be a sequence of pointed complete (also for $\nu^{\delta}$-complete) Riemannian three-manifolds having $\nu^{\delta}(p_{i})\geq \nu_{0}>0$, $\|Ric\|_{L^{s}_{g}}\leq \Lambda_{0}$ ($s>3/2$). Suppose there is a sequence of points $\{q_{i}\}$ such that $dist_{g_{i}}(q_{i},p_{i})\leq d_{0}$ and $\nu_{g_{i}}^{\delta}(q_{i})\rightarrow 0$. Then there is a subsequence converging in the weak $H^{2,s}$-topology\footnote{The notion of weak convergence of Riemannian is standard in the literature and can be stated as follows. {\it A pointed sequence $(M_{i},g_{i},p_{i})$ converges in the weak $H^{2,p}$ topology to the Riemannian space $(\bar{M},\bar{g},\bar{p})$ iff for every $\Gamma>0$ there is $i_{0}(\Gamma)$ such that for any $i>i_{0}$ there is a diffeomorphism $\varphi_{i}:B_{\bar{g}}(p,\Gamma)\cap (\bar{M}\setminus B_{\bar{g}}(\partial \bar{M},1/\Gamma))\rightarrow B_{g_{i}}(p_{i},\Gamma)\cap (M_{i}\setminus B_{g_{i}}(\partial M_{i},1/\Gamma))$ such that $\varphi^{*}g_{i}$ converges weekly in $H^{2,p}$ (with respect to the inner product defined by $\bar{g}$) to $\bar{g}$ over the region $B_{\bar{g}}(p,\Gamma)\cap (\bar{M}\setminus B_{\bar{g}}(\partial \bar{M},1/\Gamma))$}.} 
 to a $\nu^{\delta}$-complete Riemannian manifold $(\bar{\Sigma},\bar{g},p)$.} 

\vs
\n For a discussion of this and related results see \cite{Anderson}. Despite of this relevant theorem, the global geometry of $\nu^{\delta}$-complete limits may be indeed rather complicated. This represents a serious obstacle in geometric theories where the relevant quantities are $L^{s}_{g}$-norms of the Ricci curvature. The purpose of this article is to explain a series of results and conjectures showing how this deficiency is avoided granted (as the two examples above crudely show) a priori uniform lower bounds on the scalar curvature. This scenario is well suited to be applied to the Einstein CMC flow, where from the energy constraint one knows that the scalar curvature of the evolving three-metric is uniformly bounded below by minus the square of the (constant) mean curvature. The validity of the conjectures and problems posed here, would allow an improvement on the functional analysis in the space of CMC states, with potential applications in the study of singularities.  

The different sections of the article are organized as follows. In Section \ref{S2} we introduce the statements of the conjectures that will be raised.  As it turns out, some parts of them have been recently proved \cite{ReirisII} using a novel technique taken from the theory of stable minimal surfaces. We will explain the main elements of this technique and its use in the present context in Section \ref{SS2.1}. This should provide some support for the validity of the Conjectures. Despite of this, it seems unlikely that without new ideas the conjectures would be proved in their totality using minimal surface techniques only.  In Section \ref{SS2.2} we present some arguments of why a suitable notion of scalar curvature capacity (to be found) would play an essential role.  In Sections \ref{S3} and \ref{SS3.0.1} we give a brief discussion of the Constant Mean Curvature gauge from an intrinsic viewpoint using Weyl fields and Bel-Robinson energies. In Section \ref{SSS3.0.2} we prove a local equivalence between Sobolev-type norms and Bel-Robinson-type norms. This result is independent on the validity of the conjectures. We illustrate how to use this equivalence in Section \ref{SS4.3}. Applications of the conjectures are found in Sections \ref{SS4.2} and \ref{SSS4.2.1}. In particular in Section \ref{SSS4.2.1} it is shown how the conjectures (whether valid) can be applied to prove a continuity criteria for the Einstein CMC flow in terms of the $L^{4}_{g}$-norm of the space-time curvature. Proofs which require the validity of the conjecture will be headed by {\it (U.C)} (Up to the Conjectures). 

\vs
{\sf Notation.} We will denote by $H^{i,s}_{\{x\}}$ the standard $L^{s}$-Sobolev space (with its respective Sobolev norm) where the derivatives, the norms and the volume measure used in the definition are set with respect to the coordinate system $\{x\}$ (or the flat metric induced by $\{x\}$). $H^{i,s}_{g}$ is the Sobolev space (with its respective Sobolev norm) where the derivatives, the norms and the volume measure in the definition are set with respect to $g$. For instance, for a tensor $U$, we would have
\ben
\|U\|_{H^{i,s}_{g}(\Omega)}^{s}=\int_{\Omega} \sum_{j=1}^{j=i} |\nabla^{j} U|_{g}^{s} dv_{g}.
\een

{\center \section{The conjectures.}\label{S2}}

\subsection{\sf The statements.}

\begin{Conjecture}\label{C1} Let $(\Sigma,g)$ be a compact Riemannian three-manifold with $R\geq R_{0}$ and let $p$ be a number greater than $3/2$. Then

\begin{enumerate}
\item If $R_{0}>0$ then, for any $q\in \Sigma$, $1/\nu^{\delta}(q)$ is controlled from above by an upper bound on $\|Ric\|_{L^{p}_{g}}$ and $1/\nu^{\delta}(o)$. Therefore $1/Vol_{g}(\Sigma)$ is controlled from above and $Vol_{g}(\Sigma)$ from below by them too. 
\item If $R_{0}=0$ then, for any $q\in \Sigma$, $1/\nu^{\delta}(q)$ is controlled from above by an upper bound on $\|Ric\|_{L^{p}_{g}}$, $1/\nu^{\delta}(o)$ and $Vol_{g}(\Sigma)$. 
\item If $R_{0}<0$ then, for any $q\in \Sigma$, $1/\nu^{\delta}(q)$ is controlled from above by an upper bound on $\|Ric\|_{L^{p}_{g}}$, $1/\nu^{\delta}(o)$ and $dist_{g}(q,o)$.
\end{enumerate}
\end{Conjecture}

\n Note that if a statement is valid when $R_{0}<0$ then it is also valid when $R_{0}=0$ and when $R_{0}>0$. If a statement is valid when $R_{0}=0$ then it is also valid when $R_{0}>0$. Because of this one has the following {\it potential} corollary.

\begin{Corollary} (U.C) Let ${\mathcal{M}}_{R_{0},\nu_{0},\Lambda_{0}}$ be the space of complete Riemannian pointed three-manifolds $\{(\Sigma,g,o)\}$  with scalar curvature $R\geq R_{0}$, $\|Ric\|_{L^{p}_{g}(\Sigma)}\leq \Lambda_{0}$ ($p>3/2$) and $\nu^{\delta}(o)\geq \nu_{0}>0$. 
Then
\begin{enumerate}
\item The sub-family of spaces with diameter uniformly bounded above by $D_{0}$ is precompact in the weak $H^{2.p}$-topology.
\item Suppose that $R_{0}\geq 0$. Then the subfamily of spaces with $Vol_{g}(\Sigma)\leq V_{0}$ is precompact in the weak $H^{2,p}$-topology.
\end{enumerate}
\end{Corollary}   

\n In basic terms this says that there are a finite set of harmonic charts $\{\{x_{1},x_{2},x_{3}\}_{j},j=1,\ldots,m\}$ with transition functions $x_{i,k'}(x_{i,k})$ controlled in $H^{3,p}_{x_{i,k}}$, over which the metric entrances $g_{ij}$ are controlled in $H^{2,p}_{x_{i,k}}$. It is in this sense that one says that the geometry is weakly controlled in  $H^{2}$. Naturally, the limit process has to be necessarily in the weak $H^{2,p}$-topology. 

We can conjecture more relations between volume, the signature of the scalar curvature and the diameter. We synthesize them in the following.
\begin{Conjecture}\label{C2} Let $(\Sigma,g)$ be a compact three-manifold with $R\geq R_{0}$ and $\|Ric\|_{L^{p}_{g}(\Sigma)}\leq \Lambda_{0}$, and let $p$ be a number greater than $3/2$.
\begin{enumerate}
\item Suppose $R_{0}>0$. Then, for every $\epsilon>0$ there is $\Gamma(\epsilon,\Lambda_{0},R_{0})>0$ such that if $\nu^{\delta}(\Sigma)\leq \Gamma$ then $Vol_{g}(\Sigma)\leq \epsilon$. Moreover there is $D_{0}(\Lambda_{0},R_{0})$ such that $Diam_{g}(\Sigma)\leq D_{0}$.
\item Suppose $R_{0}<0$. Let $D_{0}$ be a number greater than zero. Then, for every $\epsilon>0$ there is $\Gamma(D_{0},\Lambda_{0},\epsilon,R_{0})>0$ such that if $\nu^{\delta}(\Sigma)\leq \Gamma$ and $diam(\Sigma)\leq D_{0}$ then $Vol_{g}(\Sigma)\leq \epsilon$. 

\item (from item 2) Therefore, given any $V_{0}>0$ there is $\Gamma(V_{0},D_{0},\Lambda_{0},R_{0})$ such that if $\nu^{\delta}(\Sigma)\leq \Gamma$ and $Vol_{g}(\Sigma)\geq V_{0}$ then $Diam_{g}(\Sigma)\geq D_{0}$. 

\item (from item 2) Therefore there is $\Gamma(D_{0},V_{0},\Lambda_{0},R_{0})>0$ such that if $Vol_{g}(\Sigma)\geq V_{0}$ and $diam_{g}(\Sigma)\leq D_{0}$, then there is a point $o$ for which $\nu^{\delta}(o)\geq \Gamma$. 
\end{enumerate}
\end{Conjecture}   

\n Combining {\it item 4} in Conjecture \ref{C2} with the {\it item 3} in Conjecture \ref{C1} we would get that  {\it there is $\Gamma(D_{0},V_{0},\Lambda_{0},R_{0})>0$ such that if $Vol_{g}(\Sigma)\geq V_{0}$ and $diam_{g}(\Sigma)\leq D_{0}$, then $\nu^{\delta}(\Sigma)\geq \Gamma$.} 

The following statement which is basically a restatement of Conjecture \ref{C1} when there is a priori control on the (local) curvature radius, may be of much greater utility in the study of the local geometry. Recall that, given a fixed $\Gamma>0$, the $L^{p}_{g}$-curvature radius $r_{C}^{\Gamma, p}(o)$ at a point $o$ ($p>3/2$) is defined as the supremum of the radius $r>0$ such that 
\ben
r^{2p-3}\int_{B(o,r)}|Ric|^{p}dv_{g}\leq \Gamma.
\een

\begin{Conjecture}\label{C3} Let $(\Sigma,g)$ be a complete, compact or not, Riemannian three-manifold with $R\geq R_{0}$ and let $p$ be a number greater than $3/2$.

\begin{enumerate}
\item If $R_{0}>0$ then, for any $q\in \sigma$, $1/\nu^{\delta}(q)$ is controlled from above by an upper bound on $1/r^{\Gamma,p}_{C}(\Sigma)$ and $1/\nu^{\delta}(o)$.  
\item If $R_{0}=0$ then, for any $q\in B(o,r/2)$,  $1/\nu^{\delta}(q)$ is controlled from above by an upper bound on $r$, $1/r^{\Gamma,p}_{C}(\Sigma)$, $1/\nu^{\delta}(o)$ and $Vol_{g}(B(o,r))$. 
\item If $R_{0}<0$ then, for any $q\in \Sigma$, $1/\nu^{\delta}(q)$, is controlled from above by an upper bound on $r$, $1/r^{\Gamma,p}_{C}(\Sigma)$, $1/\nu^{\delta}(o)$ and $dist_{g}(q,p)$.
\end{enumerate}
\end{Conjecture}

\n Conjecture \ref{C1}, if valid, would be in general useful to be applied in global problems, while Conjecture \ref{C3}, if valid, would be more applicable in the study of local phenomena. 

In the sections that follow we will omit writing the upper index $\delta$ and $\Gamma$ in $\nu^{\delta}$ and $r_{C}^{\Gamma}$ respectively.

\subsection{\sf Stable minimal tubes and positive scalar curvature.}\label{SS2.1}
The {\it item 1} in Conjecture \ref{C1} has been recently proved in \cite{ReirisII}. The proof is based on a novel technique that uses {\it size relations} of stable minimal tubes. Although the proof is technically involved the overall idea is simple to describe we will explain it in what follows. Let us introduce first the {\it size relations} of stable minimal tubes. 

Consider a Riemannian surface $(\Su,h)$ diffeomorphic to the tube $[-1,1]\times S^{1}$. Suppose that the stability inequality 
\be\label{II}
\int_{\Su} |\nabla f|^{2}+\kappa f^{2}dA\geq 0,
\ee

\n holds, where $f$ vanishes on the boundary of $\Su$. $\kappa$ is the Gaussian curvature of $\Su$. Such inequality arises for instance if $\Su$ is a stable minimal tube inside a three-manifold with non-negative scalar curvature. Consider now a smooth loop ${\mathcal{L}}$ embedded in $\Su$ and isotopic to the loops ${\mathcal{L}}_{1}$ and ${\mathcal{L}}_{2}$ that form the boundary of $\Su$. Let $L_{1}$ and $L_{2}$ be the distances from ${\mathcal{L}}$ to ${\mathcal{L}}_{1}$ and ${\mathcal{L}}_{2}$ respectively and let $A_{1}$ be the area of the set of points at a distance less or equal than $L_{1}$ form ${\mathcal{L}}$ in the component that contains ${\mathcal{L}}_{1}$ (and similarly for $A_{2}$). Finally let $l=length({\mathcal{L}})$, $l_{1}=length({\mathcal{L}}_{1})$ and $l_{2}=length({\mathcal{L}}_{2})$. It follows from a slight modification of a result of \cite{Castillon} (see \cite{ReirisMT} for related results in the context of this article), using the stability inequality (\ref{II}), that the following size relation holds
\be\label{CE}
2l(\frac{1}{L_{1}}+\frac{1}{L_{2}})\geq \frac{A_{1}}{L_{1}^{2}}+\frac{A_{2}}{L_{2}^{2}}.
\ee

\n Note that the expression is scale invariant. A direct consequence of this geometric relation is: {\it there are no sequences $\{\Su_{i}\}$ of stable minimal tubes satisfying
\begin{enumerate}
\item $l_{1,i}\rightarrow 0$,
\item $diam(\Su_{i})\leq L_{0}$,
\item $Area(B({\mathcal{L}}_{1,i},dist({\mathcal{L}}_{1,i},{\mathcal{L}}_{2,i})))\geq A_{0}>0$,

\item The pointed scaled spaces $(\Su_{i},\frac{1}{l_{1,i}^{2}}h_{i},p_{i})$, where $p_{i}\in {\mathcal{L}}_{1,i}$, converge (in $C^{1,\alpha}$) to the flat tube $[0,\infty)\times S^{1}$.
\end{enumerate} 
}

\n To see the claim (we will follow the terminology above) take any increasing and diverging sequence $\{d_{i}\}$ and consider the loop $\bar{{\mathcal{L}}}_{i}$ at a distance equal to $d_{i}$ from ${\mathcal{L}}_{1,i}$ in the metric $\frac{1}{l_{1,i}^{2}}h_{i}$. Suppose too that $\{d_{i}\}$ was chosen in such a way when $h_{i}$ is scaled as in {\it item 4} the region enclosed by ${\mathcal{L}}_{1,i}$ and $\bar{{\mathcal{L}}}_{i}$ is closer and closer (globally) to the tube $[0,d_{i}]\times S^{1}$. Now if we chose as ${\mathcal{L}}_{i}$ the loop at a distance $d_{i}/2$ from ${\mathcal{L}}_{1,i}$, it follows from the scale invariance of the expression (\ref{CE}) that $l_{i}(1/L_{1,i}+1/L_{2,i})\rightarrow 0$. But the right hand side of (\ref{CE}) is bounded below by 
\ben
\frac{A_{1,i}}{L_{1,i}^{2}}+\frac{A_{2,i}}{L_{2,i}^{2}}\geq \frac{A_{1,i}+A_{2,i}}{L_{0}^{2}}\geq \frac{A_{0}}{L_{0}^{2}}.
\een

\n which is a contradiction. 

The idea to prove {\it item 1} is the following. Proceed by contradiction and assume there is a sequence of spaces $\{(\Sigma_{i},g_{i},o_{i})\}$ such that $\|Ric\|_{L^{p}_{g_{i}}(\Sigma_{i})}\leq \Lambda_{0}$ and $R_{g_{i}}\geq R_{0}>0$, $\nu(o_{i})\geq \nu_{0}>0$ but $\nu(\Sigma_{i})\rightarrow 0$. Note that because $\|Ric\|_{L^{p}_{g_{i}}(\Sigma_{i})}\leq \Lambda_{0}$ and $R_{g_{i}}\geq R_{0}>0$ then $Vol_{g_{i}}(\Sigma_{i})\leq R_{0}^{-p}\Lambda_{0}^{p}$. Because of this and Theorem {\it Convergence II}, a subsequence has a limit space $(\tilde{\Sigma},\tilde{g},\tilde{o})$ with $\nu(\tilde{\Sigma})=0$. Then it is proved, using the theory of collapse under $L^{p}_{g}$-curvature bounds in dimension three, that if $\epsilon_{0}$ is small enough, then the topology of the $\epsilon_{0}$-thin region $\tilde{\Sigma}_{\epsilon_{0}}$ (namely the set of points $q$ in $\tilde{\Sigma}$ with $\nu(q)\leq \epsilon_{0}$) allows to construct a sequence of stable minimal tubes $S_{i}$ in $\tilde{\Sigma}$ such that: one of the boundaries ${\mathcal{L}}_{1,i}$ remains fixed while the size of the other ${\mathcal{L}}_{2,i}$ goes to zero. Because ${\mathcal{L}}_{1,i}$ remains fixed it is easily proved that $A(S_{i})\geq A_{0}>0$. On the other hand a well known argument due to Fischer-Colbrie shows that if $R_{0}>0$ then $diam(S_{i})\leq d_{0}$. Thus we have satisfied the three first {\it items} in the list above. The {\it item 4} can be seen to follow from a scaling argument and the a priori $L^{p}_{g}$ curvature bound. A contradiction is thus reached. It is straightforward to put this argument at work in Examples \ref{EXC1} and \ref{EXC2}.    

\subsection{\sf Capacity and scalar curvature.}\label{SS2.2}

In this informal section we will discuss further alternatives to approach the Conjectures. 

As it turns out, further progress in {\it items 2} and {\it 3} of Conjecture \ref{C1} can be carried out using minimal surface techniques. However without new ideas it is unlikely that a complete proof can be achieved. Indeed it seems that {\it item 1} can be used to reduce {\it items 2} and {\it 3} into a problem of scalar curvature capacity. As it is well known, in general it is possible to lower the value of the scalar curvature inside a definite region by suitably deforming the metric inside the region\cite{Lohkamp} (with great freedom), however certain capacity must exists inside if one is willing to increase it. Specifying the nature of such capacity is a difficult task and in a certain sense depends very much on the a priori question of what precisely is that such capacity is intended to measure. Leaving aside these considerations we will make the claim (we will leave it as a problem in fact) that such capacity exists on $\epsilon$-thin regions under $L^{p}_{g}$-curvature bounds, allowing to rise the value of the given scalar curvature by any amount if $\epsilon$ is small enough. A precise formulation is in the following problem.  

\begin{Problem}\label{P1} (Prove or disprove) Let $\{(\Sigma,g,o)\}$ be a complete Riemannian manifolds with 
\begin{enumerate}
\item $\|Ric\|_{L^{p}_{g}(\Sigma)}\leq \Lambda_{0}$,
\item $\nu_{g}(o)\geq \nu_{0}$,
\item $Vol_{g}(\Sigma)\leq V_{0}$.
\end{enumerate}

\n Then, there exist $\bar{\nu}$, $\bar{\Lambda}_{0}$, and $f:\field{R}^{+}\rightarrow \field{R}^{+}$ with $f(x)\rightarrow 0$ when $x\rightarrow 0$, all depending on $\Lambda_{0}$, $\nu_{0}$ and $V_{0}$ such that for any positive number $\tilde{R}$ there are $\epsilon(\nu_{0},\Lambda_{0},V_{0},\tilde{R})$ and a (deformed) metric $\bar{g}$ satisfying
\begin{enumerate}
\item $R_{\bar{g}}\geq R_{g}+\tilde{R}$ on $\Sigma_{g,\epsilon}$ and $R_{\bar{g}}\geq R_{g}$ everywhere else,
\item $\|Ric\|_{L^{p}_{\bar{g}}(\Sigma)}\leq \bar{\Lambda}_{0}$,
\item $\nu_{\bar{g}}(o)\geq \bar{\nu_{0}}$,
\item $\nu_{\bar{g}}(p)\leq f(\nu_{g}(p))$ for every $p\in \Sigma$.
\end{enumerate}
\end{Problem}     

\n Note that there is no reference to any lower bound on $R_{g}$. It is evident that if Problem \ref{P1} is correct then so is {\it item 1} in Conjecture \ref{C1}. The {\it item 2} would be also valid if we assume a further control in the total volume (as in {\it item 1}). Note too that one must assume that at some point $o$ the space is non-collapsed, otherwise the statement is clearly wrong (consider a collapsed flat three-torus).

There is a heuristic reason of why this capacity should exists in this context. From the simple fact that the volume comparison must fail one deduces that, on very thin regions and under $L^{p}_{g}$ curvature bounds, the $L^{\infty}_{g}$-norm of $Ricci$ should be big (the bigger the thiner). This large value of $|Ric|$ (in most of the thin region) should provide the necessary capacity to rise the scalar curvature. 

Conformal deformations of the metric work in some cases but it is unlikely they would provide (if valid) a solution to Problem \ref{P1}. However it is worth to note that conformal geometry gives an interesting avenue to prove (at least) {\it item 2} in Conjecture \ref{C1}. Let us illustrate (rather informally) how this could be done through a simple example. Consider a conformal metric $g=\phi^{4}g_{F}$ on $T^{2}\times \field{R}$, where $g_{F}$ is the flat product metric, and suppose that $\phi(p)\rightarrow 0$ as $p\rightarrow \infty$. It is easy to see that $R$ cannot be non-negative. Indeed if it were then we would have
\ben 
R\phi^{5}=-8\Delta_{g_{F}} \phi.
\een

\n So we have $\Delta \phi\leq 0$ and $\phi>0$. An easily adapted version of a theorem of Yau and Cheng \cite{Cheng-Yau} shows that if the volume of balls $B_{g}(o,r)$ grows at least quadratically when $r\rightarrow\infty$ then there does not exist positive superharmonic functions decaying to zero, a situation that directly applies here. This shows that cusps with this kind of geometry and $R_{g}\geq 0$ do not exists. But what is interesting is that the argument underlying this example may work in a much broader context. Suppose that $(\tilde{\Sigma},\tilde{g},\tilde{o})$ is the limit of a sequence of pointed spaces $\{(\Sigma_{i},g_{i},o_{i})\}$ with $R_{g_{i}}\geq 0$ and uniformly bounded $L^{p}_{g_{i}}$-curvature. Then consider the metric $g_{S}(p)=(1/\nu(p))^{2}\tilde{g}(p)$ (discard for the moment regularity issues). We have the following facts (we are being informal)
\begin{enumerate}
\item 
the metric $g_{S}$ around a point $q$ becomes flatter and flatter as $q\rightarrow \infty$,
\item 
if $q$ is sufficiently far away then there will be one or two fibers collapsed at $q$ and therefore the geometry would look (when looked from far away) as one or two dimensional. 
\end{enumerate}   

\n We also have (make $\phi^{2}(p)=\nu(p)$) 
\ben
0\geq -R_{\tilde{g}}\phi^{5}= 8\Delta_{g_{S}} \phi -R_{g_{S}}\phi,
\een

\n If $R_{g_{S}}(q)$ decays to zero sufficiently fast as $q\rightarrow \infty$ then it would be possible to 
\begin{enumerate}
\item 
find a conformal transformation of $g_{S}$, $g_{Sc}=\varphi^{4}g_{S}$ with $\varphi\geq c>0$ and $R_{g_{Sc}}=0$,
\item
for which the volume $Vol_{g_{Sc}}(B_{g_{Sc}}(o,r))$ would grow at most quadratically in $r$ as $r\rightarrow \infty$,
\end{enumerate}

\n Making $\xi=\phi/\varphi$ we would have,
\ben
\Delta_{g_{Sc}} \xi\leq 0,
\een

\n and the Cheng-Yau theorem would give a contradiction. It is indeed possible to design  conditions on Ricci to guarantee that the procedure described above can be carried out. However a proof of {\it item 2} in Conjecture \ref{C1} based only on this kind of arguments seems to be at least difficult.

{\center \section{Elements of the Einstein flow.}}
\subsection{\sf The Constant Mean Curvature gauge.}\label{S3}

We will not enter into describing a theory of the Constant Mean Curvature (CMC) gauge in (vacuum) General Relativity. A detailed description can be found found in \cite{ReirisCMC} and references therein. We will take a practical viewpoint and think the vacuum Einstein equations as a flow. For the sake of simplicity we will consider space-times with compact Cauchy hypersurfaces. Such space-times are usually called {\it Cosmological}. In later sections we will explore potential applications of the conjectures that were raised before. We will do that in the context of Cosmological space-times, despite of the fact that every statement has an analogous one in the context of {\it maximal} asymptotically flat (AF) space-times. It should be remarked that, from the point of view of the conjectures, the maximal flow would have advantages in comparison to the cosmological scenario. Indeed the condition $R\geq 0$ assures, if the conjectures are correct, that it is not necessary to control the distance function but instead an a priori control on the volume, which is a much simpler invariant, is enough. In fact, if we choose the zero shift (see below) in the Einstein maximal flow, the volume of a fixed region $\Omega$ is preserved during the evolution and therefore automatically controlled. 

Formally, an Einstein CMC flow over a compact three-manifold $\Sigma$ and over a range of CMC-times $[k_{0},k_{1}]$ is a flow $k\rightarrow (g,K,X)$, where $g$ is a Riemannian three-metric, $K$ is a two-tensor and $X$ a vector field, the so called {\it shift vector}. The space-time associated with such flow is given by  $({\bf M},{\bf g})$ where ${\bf M}=[k_{0},k_{1}]\times \Sigma$ and the metric is given by
\ben
\phi^{*}{\bf g}=-(N^{2}-|X|^{2})dt^{2}+dt\otimes X^{*}+X^{*}\otimes dt + g.
\een

\noindent In the formula above $g$ was extended to ${\bf M}$ to be zero along
$\partial_{t}$ (i.e. zero when one of the entrances is $\partial_{t}$) and $X^{*}_{a}=X^{b}g_{ab}$. $N$ is the so called {\it lapse} and satisfies the fundamental {\it lapse equation}
\be\label{LE}
-\Delta N + |K|^{2}N=1.
\ee

\n We are making $t=k$. The two-tensor $K(k)$ is the second fundamental form of the slice $\{k\}\times \Sigma$ with respect to ${\bf g}$ and $k$ is the mean curvature, i.e. $k=tr_{g(k)}K(k)$.  The Einstein equations split into two groups, the {\it constraint equations}
\be\label{1.1}
R=|K|^{2}-k^{2},
\ee
\be\label{2.1}
div\ K=dk,
\ee

\n and the {\it dynamical equations} or {\it Hamilton-Jacobi equations}
\be\label{3.1}
\dot{g}=-2NK+{\mathcal{L}}_{X}g,
\ee
\be\label{4.1}
\dot{K}=-\nabla\nabla N+N(Ric +kK-2KK)+{\mathcal{L}}_{X}K.
\ee  

In practice we will assume that $k_{0}<k_{1}<0$. Including the case $k=0$ would bring some difficulties, that although mainly of a technical nature, they would shadow and enlarge the presentation. 

\subsection{\sf Weyl fields and the Bel-Robinson energies as a theoretical framework.}\label{SS3.0.1}

There is a theoretical interesting approach to the Einstein equations through Weyl fields which has given remarkable results (see for instance \cite{Ch-K}). The advantage of working with Weyl fields is not only esthetic, it provides a comprehensive display of the algebraic and evolutionary properties of curvatures and their derivatives and ultimately amount to have full control of the metric properties of the space-time. In vacuum, the situation we are in in this article, the most basic Weyl field is the space-time Riemann tensor ${\bf W}_{0}={\bf Rm}$. Higher order Weyl fields are constructed as ${\bf W}_{i}=\bn^{i}_{T} {\bf Rm}$, $i=1,2,\ldots$, where $T$ is the normal vector field to the CMC foliation of the space-time. The viewpoint of describing curvatures through Weyl field is the one we will adopt here and used in the applications of later sections. For this reason we need to introduce below and in the shortest way the basic elements of the framework \cite{Ch-K}. 

A Weyl field ${\bf W}={\bf W}_{abcd}$ is a traceless $(4,0)$ space-time 
tensor field having the symmetries of the curvature tensor ${\bf Rm}$. Given a Weyl tensor ${\bf W}$ define the left and
right duals by $^{*}{\bf W}_{abcd}=\frac{1}{2}\epsilon_{ablm}{\bf W}^{lm}_{\ \ cd}$ and ${\bf W}^{*}_{abcd}={\bf W}_{ab}^{\ \ lm}
\frac{1}{2}\epsilon_{lmcd}$ respectively. It is $^{*}{\bf W}={\bf W}^{*}$ and $^{*}(^{*}{\bf W})=-{\bf W}$. Define the current ${\bf J}$ 
and its dual ${\bf J}^{*}$ by $\bn^{a}{\bf W}_{abcd}={\bf J}_{abc}$, $\bn^{a}{\bf W}^{*}_{\ abcd}={\bf J}^{*}_{abc}$. When ${\bf W}$ is the Riemann tensor the currents ${\bf J}$ and ${\bf J}^{*}$ are zero due to the Bianchi identities. The $L^{2}$-norm with respect to the foliation will be defined through the Bel-Robinson tensor. Given a Weyl
field ${\bf W}$ define the Bel-Robinson tensor by
\ben
Q_{abcd}({\bf W})={\bf W}_{alcm}{\bf W}_{b\ d}^{\ l\ m}+{\bf W}_{alcm}^{*}{\bf W^{*}}_{b\ d}^{\ l\ m}.
\een
    
\noindent The Bel-Robinson tensor is symmetric and traceless in all pair of indices and for any pair of timelike vectors $T_{1}$ and $T_{2}$,
the quantity $Q_{T_{1}T_{2}T_{3}T_{4}}=Q(T_{1},T_{1},T_{2},T_{2})$ is non-negative. The electric and magnetic components of ${\bf W}$ are defined by
\ben
E_{ab}={\bf W}_{acbd}T^{c}T^{d},
\een
\ben
B_{ab}=^{*}{\bf W}_{acbd}T^{c}T^{d}.
\een

\noindent The tensor fields $E$ and $B$ are symmetric, traceless and null in the $T$ direction. It is also the case that ${\bf W}$ can be reconstructed from them \cite{Ch-K}. If ${\bf W}$ is the Riemann tensor in a vacuum solution we have
\begin{equation}\label{de}
E_{ab}=Ric_{ab}+kK_{ab}-K_{a}^{\ c}K^{c}_{\ b},
\end{equation}
\begin{equation}\label{de2}
\epsilon_{ab}^{\ \ l}B_{lc}=\nabla_{a}K_{bc}-\nabla_{b}K_{ac}.
\end{equation}

\n Note that equation (\ref{de2}) together with equation (\ref{2.1}) and the CMC condition of $k$ being constant implies the elliptic system for $K$
\be\label{EK1}
div K=0,
\ee
\be\label{EK2}
curl K=B,
\ee

\n The components of a Weyl field with respect to the CMC foliation are given by ($i,j,k,l$ are spatial indices) 
\ben
{\bf W}_{ijkT}=-\epsilon_{ij}^{\ \ m}B_{mk},\ ^{*}{\bf W}_{ijkT}=\epsilon_{ij}^{\ \ m}E_{mk},
\een
\ben
{\bf W}_{ijkl}=\epsilon_{ijm}\epsilon_{kln}E^{mn},\ ^{*}{\bf W}_{ijkl}=\epsilon_{ijm}\epsilon_{kln}B^{mn}.
\een

\noindent We also have 
\ben
Q_{TTTT}=|E|^{2}+|B|^{2},
\een 
\ben
Q_{iTTT}=2(E\wedge B)_{i},
\een
\ben
Q_{ijTT}=-(E\times E)_{ij}-(B\times B)_{ij}+\frac{1}{3}(|E|^{2}+|B|^{2})g_{ij}.
\een

\n The operations $\times$ and $\wedge$ are provided explicitly later. The divergence of the Bel-Robinson tensor is
\ben
\begin{split}
\bn^{a}Q({\bf W})_{abcd}=&{\bf W}_{b\ d}^{\ m\ n}{\bf J}({\bf W})_{mcn}+{\bf W}_{b\ c}^{\ m\ n}{\bf J}({\bf W})_{mdn}\\
&+^{*}{\bf W}_{b\ d}^{m\ n}{\bf J}^{*}({\bf W})_{mcn}+^{*}{\bf W}_{b\ c}^{m\ n}{\bf J}^{*}(W)_{mcn}.
\end{split}
\een

\n We have therefore
\ben
\bn^{\alpha}Q({\bf W})_{\alpha TTT}=2E^{ij}({\bf W}){\bf J}({\bf W})_{iTj}+2B^{ij}{\bf J}^{*}({\bf W})_{iTj}.
\een

\noindent From that we get the {\it Gauss equation} which will be used later in the article
\be\label{Gausseq}
\begin{split}
\dot{Q}({\bf W})=&-\int_{\Sigma}2NE^{ij}({\bf W}){\bf J}({\bf W})_{iTj}+2NB^{ij}({\bf W}){\bf J}^{*}({\bf W})_{iTj}\\
&+3NQ_{abTT}\dt^{ab}dv_{g}.
\end{split}
\ee

\noindent $\dt_{ab}=\bn_{a}T_{b}$ is the {\it deformation tensor} and plays a fundamental role in the space-time 
tensor algebra. In components it is
\ben
\dt_{ij}=-K_{ij}, \ \ \ \dt_{iT}=0,
\een
\ben
\dt_{Ti}=\frac{\nabla_{i}N}{N}, \ \ \ \dt_{TT}=0.
\een

The next equations are essential when it comes to get elliptic estimates of Weyl fields.
\begin{equation}\label{eq1}
div E({\bf W})_{a}=(K\wedge B({\bf W}))_{a}+{\bf J}_{TaT}({\bf W}),
\end{equation}
\begin{equation}\label{eq2}
div B({\bf W})_{a}=-(K\wedge E({\bf W}))_{a}+{\bf J}^{*}_{TaT}({\bf W}),
\end{equation}
\begin{equation}\label{eq3}
curl B_{ab}({\bf W})=E(\bn_{T}{\bf W})_{ab}+\frac{3}{2}(E({\bf W})\times K)_{ab}-\frac{1}{2}kE_{ab}({\bf W})+{\bf J}_{aTb}({\bf W}),
\end{equation}
\begin{equation}\label{eq4}
curl E_{ab}({\bf W})=B(\bn_{T}{\bf W})_{ab}+\frac{3}{2}(B({\bf W})\times K)_{ab}-\frac{1}{2}kB_{ab}({\bf W})+{\bf J}^{*}_{aTb}({\bf W}).
\end{equation}

\noindent The operations $\wedge,\ \times$ and the operators $div$ and $curl$ are defined through
\ben
(A\times B)_{ab}=\epsilon_{a}^{\ cd}\epsilon_{b}^{\ ef}A_{ce}B_{df}+\frac{1}{3}(A\circ B)g_{ab}-\frac{1}{3}(tr A)(tr B)g_{ab},
\een
\ben
(A\wedge B)_{a}=\epsilon_{a}^{\ bc}A_{b}^{\ d}B_{dc},
\een
\ben
(div\ A)_{a}=\nabla_{b}A^{b}_{\ a},
\een
\ben
(curl\ A)_{ab}=\frac{1}{2}(\epsilon_{a}^{\ lm}\nabla_{l}A_{mb}+\epsilon_{b}^{\ lm}\nabla_{l}A_{ma}).
\een
  
\subsection{\sf Sobolev norms versus Bel-Robinson type norms.}\label{SSS3.0.2}

As we expressed before, the Bel-Robinson energies are interesting quantities to control de Einstein flow. 

\begin{Lemma} {\rm (Sobolev norms vs. Bel-Robinson norms)}\label{Intr1} Let $\bar{I}\geq 0$ and let $\Sigma$ is a compact
three-manifold. Then the functional (defined over states $(g,K)$ with $|k|\leq 1$) 
\ben
\|(g,K)\|_{BR}=\frac{1}{\nu(\Sigma)}+{\mathcal{V}}+\sum_{j=0}^{j=\bar{I}}Q_{j},
\een

\n where ${\mathcal{V}}=-k^{3}Vol_{g}(\Sigma)$, controls the reciprocal of the $H^{\bar{I}+2}$-harmonic radius\footnote{The $H^{i}$-{\it harmonic radius} $r^{i}_{H}(o)$ at $o$ in a $H^{i+1}$-Riemannian three-manifold $(\Sigma,{\mathcal{A}},g)$, $i\geq 2$, 
is defined as the supremum of the radius $r$ for which there is a coordinate chart $\{x\}$ covering $B(o,r)$ 
and satisfying
\ben
\frac{3}{4}\delta_{jk}\leq g_{jk}\leq \frac{4}{3}\delta_{jk},
\een
\ben
\sum_{\alpha=2}^{\alpha=i}r^{2\alpha-3}(\sum_{|I|=\alpha,j,k}\int_{B(o,r)}|\frac{\partial^{I}}{\partial x^{I}}g_{jk}|^{2}dv_{x})\leq 1,
\een

\n Thus, in a sense, the $H^{i}$-harmonic radius controls the $H^{i}$-Sobolev norm of $g_{ij}-\delta_{ij}$ in suitable harmonic coordinates.} $r_{H}^{\bar{I}+2}$ and the $H^{\bar{I}+1}_{g}$-norm of $K$ from above.
\end{Lemma}

\n There is a difference between the case $\bar{I}=0$ and $\bar{I}>0$. The first case follows from intrinsic estimates which are independent on $\nu(\Sigma)$. These estimates are given by
\be\label{EE1}
\|\nabla K\|^{2}_{L^{2}_{g}}+\|K\|^{4}_{L^{4}_{g}}=\int_{\Sigma}|\nabla \hat{K}|^{2}+|\hat{K}|^{4}dv_{g}\leq C(|k|{\mathcal{V}}+Q_{0}),
\ee
\be\label{EE2}
\|Ric\|^{2}_{L^{2}_{g}}=\int_{\Sigma}|Ric|^{2}dv_{g}\leq C(|k|{\mathcal{V}}+Q_{0}).
\ee

\n where $C$ is numeric \cite{ReirisCMC}. When $\bar{I}>0$ however, elliptic estimates are required and therefore the dependence on $\nu(\Sigma)$ becomes necessary. It is not difficult indeed to guess a proof of the Lemma \ref{Intr1} from the case $\bar{I}=0$ and the elliptic system (\ref{eq1})-(\ref{eq4}) (see \cite{ReirisCMC}).   

It is interesting that this Lemma has a local version showing that it is not necessary to have global control on the $L^{2}_{g}$-norms of $E$ and $B$ to have local control of $(g,K)$ in $H^{2}\times H^{1}$ (locally and in harmonic coordinates). Unfortunately when only local information on $E$ and $B$ is provided we have no longer the intrinsic estimates (\ref{EE1}) and (\ref{EE2}) and therefore the $\bar{I}=0$ case already requires a priori control on $\nu(p)$, where $p$ is the point where the estimates is carried out. We will provide such local estimates in the proposition below. It will be required later in the study of velocity dominated singularities. The Lemma is independent of the Conjectures.  As usual local information should be provided in the form of radiuses. We have already defined the $L^{2}_{g}$-curvature radius in Section \ref{S2}. More generally, suppose that $U$ is a tensor such that $|U|$ scales as the reciprocal of the distance square. Then define the $L^{2}_{g}$- {\it U - radius}, $r_{U}(p)$, at the point $p$, as the supremum of the the radius $r>0$ such that
\ben
r\int_{B_{g}(p,r)}|U|^{2}dv_{g}\leq \Gamma,
\een

\n where $\Gamma$, as in the curvature radius, is set fixed. In this way one can consider the radiuses $r_{E}(p)$, $r_{B}(p)$, $r_{K}(p)$ and $r_{\nabla K}(p)$ corresponding, respectively, to $U=E$, $U=B$, $U=K\circ K$ and $U=\nabla K$. The {\it Bel-Robinson} radius $r_{Q_{0}}(p)$ is defined by replacing $|U|^{2}$ in the equation before by $|E|^{2}+|B|^{2}$.

\begin{Lemma}\label{LF} Let $\Sigma$ be a compact three-manifold and let $(g,K)$ be a CMC state with $|k|\leq 1$. Let $r$ be a fixed radius and let $p$ be any point in $\Sigma$. Suppose that for some constants $c_{1}>0$ and $c_{2}>0$ we have 
\be\label{FR}
r_{Q_{0}}(q)\geq c_{1} dist(q,\partial B(p,r)),
\ee
\be\label{FV}
\nu(q)\geq c_{2} dist(q,\partial B(p,r)),
\ee

\n for every $q$ in $B(p,r)$. Then there is $c_{3}(c_{1},c_{2},r)>0$ such that $r_{K}(q)\geq c_{3} dist(q,\partial B(p,r))$ and similarly for $r_{C}(q)$. 
\end{Lemma}  

\n {\it Proof:} 

Suppose, by contradiction, that for every integer $i>1$ there is a state $(g_{i},K_{i})$ and $p_{i}\in \Sigma$ for which (\ref{FR}) and (\ref{FV}) are valid but for some $q_{i}$ in $B_{g_{i}}(p_{i},r)$ it is
(denote $B_{g_{i}}(p_{i},r)=B_{i}(r)$, $dist_{g_{i}}(q,\partial B_{i}(r))=d_{i}(q)$)
\ben
\frac{d_{i}(q_{i})}{i}\int_{B_{i}(q_{i},\frac{d_{i}(q_{i})}{i})}|K_{i}|^{4}dv_{g_{i}}\geq \Gamma.
\een

\n We deduce that there is $f(i)\geq i$ such that 
\ben
\sup_{q\in B_{i}(r)}\frac{d_{i}(q)}{f(i)} \int_{B_{i}(q,\frac{d_{i}(q)}{f(i)})}|K_{i}|^{4}dv_{g_{i}}=\Gamma,
\een

\n To simplify notation let $q_{i}\in int B_{i}(r)$ be a point at which equality above is reached. From this one easily shows that there is a sequence of states, denoted again by $(g_{i},K_{i})$, and an increasing function $f:\field{Z}^{+}\rightarrow \field{R}^{+}$, such that
\ben
\frac{d_{i}(q)}{f(i)} \int_{B_{i}(q,\frac{d_{i}(q)}{f(i)})}|K_{i}|^{4}dv_{g_{i}}\leq \frac{d_{i}(q_{i})}{f(i)} \int_{B_{i}(q_{i},\frac{d_{i}(q_{i})}{f(i)})}|K_{i}|^{4}dv_{g_{i}}=\Gamma,
\een

\n for every $q\in B_{i}(r)$. Scale the states $(g_{i},K_{i})$ to get a new state $(\tilde{g}_{i},\tilde{K}_{i})$ with

\n $\tilde{d}_{i}(q_{i})\equiv dist_{\tilde{g}_{i}}(q_{i},\partial B_{i}(r))=f(i)$ and therefore $\tilde{r}_{K}(q_{i})\equiv r_{K,(\tilde{g}_{i},\tilde{K}_{i})}(q_{i})=1$. Note that we have  
\ben
\tilde{r}_{K}(q)\geq \frac{\tilde{d}_{i}(q)}{f(i)}\geq \frac{\tilde{d}_{i}(q_{i})-dist_{\tilde{g}_{i}}(q_{i},q)}{f(i)}=1-\frac{dist_{\tilde{g}_{i}}(q,q_{i})}{f(i)}.
\een

\n Using this we can extract a subsequence (denoted again by $(\tilde{g}_{i},\tilde{K}_{i})$) with the following properties. For any given $d_{0}>0$ and $q$ such that $dist_{\tilde{g}_{i}}(q,q_{i})\leq d_{0}$ it is  
\begin{enumerate}
\item $\tilde{r}_{K}(q)\geq \frac{1}{2}$ if $i\geq i_{0}(d_{0})$, 

\item $r_{Q_{0}(\tilde{g}_{i},\tilde{K}_{i})}(q)\rightarrow \infty$,

\item $\nu_{\tilde{g}_{i}}(q)\rightarrow \infty$.
\end{enumerate}

\n It follows from here that for $i\geq i_{0}(d_{0})$ and $i_{0}(d_{0})$ sufficiently big then $\tilde{r}_{C,i}(q)\geq c_{4}$ if $dist_{\tilde{g}_{i}}(q,q_{i})\leq d_{0}$ and $c_{4}$ numeric. Thus we can take a convergent subsequence to a state $(\tilde{g}_{\infty},\tilde{K}_{\infty})$, with $\nu_{\tilde{g}_{\infty}}=\infty$, satisfying the elliptic system
\be\label{EB1}
Ric_{\tilde{g}_{\infty}}=\tilde{K}_{\infty}\circ \tilde{K}_{\infty},
\ee
\be\label{EB2}
curl\ \tilde{K}_{\infty}=0,
\ee
\be\label{EB3}
div\ \tilde{K}_{\infty}=0,
\ee

\n and thus being $C^{\infty}$. To see that $\tilde{K}_{\infty}\neq 0$ we note first that from the elliptic system (\ref{EK1})-(\ref{EK2}) and standard elliptic estimates we know that $\|\tilde{K}_{i}\|_{H^{1}_{\tilde{g}_{i}}(B_{\tilde{g}_{i}}(q_{i},2))}$ is uniformly bounded above. This together with the fact that $\tilde{r}_{K,i}(q_{i})=1$ implies that in the limit $\tilde{r}_{K,\infty}(q_{\infty})=1$ and thus $\tilde{K}_{\infty}\neq 0$. In the same fashion, from $\tilde{r}_{K,\infty}(q)\geq 1/2$ we can deduce $\|\tilde{K}_{\infty}\|_{L^{\infty}_{\tilde{g}_{\infty}}}=c_{5}$ where $\infty>c_{5}>0$. As we will briefly explain below, the elliptic system (\ref{EB1})-(\ref{EB3}) implies 
\be\label{Simons}
\Delta |\tilde{K}_{\infty}|^{2}=2|\nabla \tilde{K}_{\infty}|^{2}+5|\tilde{K}_{\infty}|^{4}.
\ee

\n It is easy to see that this implies a contradiction by taking a sequence of points $s_{i}$ seeking for the supremum of $|\tilde{K}_{\infty}|$ (i.e. such that $\lim |\tilde{K}_{\infty}|(s_{i})=\|\tilde{K}_{\infty}\|_{L^{\infty}_{\tilde{g}_{\infty}}}$) and using equation (\ref{Simons}). This would finish the proof of the Lemma.

We show now that a state $(g,K)$ with $k=0$ and satisfying the system 
\ben
Ric=K\circ K,\ curl\ K=0,\ div\ K=0,
\een

\n satisfies an equation of the type (\ref{Simons}). Recall that for any symmetric two tensor $K$ we have
\be\label{Simons2}
d^{\nabla*}d^{\nabla}\ K+2div^{*}div\ K=2\nabla^{*}\nabla\ K+{\mathcal{R}}(K),
\ee

\n where ${\mathcal{R}}(K)=Ric\circ K+K\circ Ric-2Rm\circ K$ and $Rm\circ K=Rm_{acbd}K^{cd}$. Also recall the formula for $Rm$ in terms of $Ric$
\ben
Rm_{ijkl}=g_{ik}Ric_{jl}+g_{jl}Ric_{ik}-g_{jk}Ric_{il}-g_{il}Ric_{jk}-\frac{1}{2}(g_{ik}g_{jl}-g_{jk}g_{il})R.
\een
\n So we have ($k=0$)
\ben
Rm\circ K=g_{ik}<Ric,K>-(Ric\circ K)_{ik}-(K\circ Ric)_{ik}+\frac{1}{2}K_{ik}R.
\een

\n Therefore 
\ben
<K,{\mathcal{R}}K>=-2<K,Ric\circ K>+\frac{1}{2}|K|^{2}R,
\een

\n and 
\ben
2<K,\nabla^{a}\nabla_{a}K>=5|K|^{4},
\een

\n Altogether we have
\ben
\Delta|K|^{2}=2|\nabla K|^{2}+5|K|^{4}.
\een\ep 

{\center \section{(Potential) applications to General Relativity.}\label{S4}}

\subsection{\sf Some simple (potential) geometric applications.}\label{SS4.1}

The following proposition gives a first simple potential application of the Conjectures to the AF manifolds of non-negative scalar curvature.  

\begin{Proposition}(U.C)  Let $(\Sigma,g)$ be a non-compact three-manifold of non-negative scalar curvature. Assume that outside a compact set ${\mathcal{K}}$ it is $\Sigma\setminus {\mathcal{K}}\approx \field{R}^{3}\setminus B(o,1)$. Suppose too that the weighted norm\footnote{We assume there is a harmonic coordinate system $\{x_{i}\}$ at infinity for which we have 
$\sup r^{-1} |g_{ij}-\delta_{ij}|+r^{-2}|\partial_{k}g_{ij}|+r^{-2+\alpha}|\partial_{k}g_{ij}(x)-\partial_{k}g_{ij}(y)|/|x-y|^{\alpha}\leq c_{0}$.}  $\|g_{ij}-\delta_{ij}\|_{C_{w}^{1,\alpha}}\leq c_{0}$ for some fixed $c_{0}$, and that $Vol_{g}({\mathcal{K}})\leq V_{0}$ and $\|Ric\|_{L^{2}_{g}(\Sigma)}\leq \Lambda_{0}$. Then
\begin{enumerate}
\item $1/\nu(\Sigma)$ is controlled from above by $\Lambda_{0},c_{0},V_{0}$. 

\item Fix $r>0$. Suppose we have a sequence of spaces $\{(\Sigma_{i},g_{i})\}$ with the same uniform bounds as in the previous item.  Then $\int_{B({\mathcal{K}},r)}R_{g_{i}}dv_{g_{i}}\rightarrow 0$ if $m(g_{i})\rightarrow 0$. 

\item Suppose that $K$ (which has a boundary diffeomorphic to $S^{2}$) is topologically different from a three ball. Then $m\geq m(c_{0},\Lambda_{0},V_{0})>0$.
\end{enumerate}
\end{Proposition}   

\n {\it Sketch of the proof:} 

	{\it Item 1}. This {\it item} would follow from the {\it item 2} in Conjecture \ref{C3}. Note that $1/\nu(p)$ is controlled from above on any point $p$ outside ${\mathcal{K}}$. 
	
	{\it Item 2}.  In this {\it item} one would proceed by contradiction and take a convergent subsequence into an AF space $(\Sigma_{\infty},g_{\infty})$ with $\int_{\Sigma_{\infty}}R_{g_{\infty}}dv_{g_{\infty}}>0$ but zero ADM mas, which is a contradiction.   

	{\it Item 3}. This follows the same idea as in the previous {\it item} but this time the limit would have zero mass and non-trivial topology, which is a contradiction.\ep

\subsection{\sf (Potential) applications to the Cauchy problem and the regularity of the space-time.}\label{SS4.2}

\subsubsection{\sf The space-time curvature and the space-time volume radius.}

\begin{Proposition}(U.C) Let $(\Sigma,(g,K))$ be a CMC state with $|k|\leq 1$, over a compact three-manifold of arbitrary topology. Then the inverse $1/N$ of the lapse $N$ is controlled from above by an upper bound on $diam(\Sigma)$, $1/Vol_{g}(\Sigma)$ and $Q_{0}$.
\end{Proposition}

\n {\it Sketch of the proof:} 

The proof would follow from two facts. The first is that, from inequality (\ref{EE2}) and Conjecture $\ref{C3}$ ({\it item 2}) it follows that the reciprocal of the volume radius of the total manifold is controlled from above by an upper bound on $diam(\Sigma),\ 1/Vol_{g}(\Sigma)$ and $Q_{0}$. Thus this gives control of $(g,K)$ in $H^{2}\times H^{1}$ (in harmonic coordinates). We can use therefore standard elliptic analysis (we will do that implicitly next). The second fact follows from analyzing the lapse equation. Multiply the lapse equation 
\ben 
-\Delta N+|K|^{2}N=1,
\een

\n by $\frac{1}{N^{3}}$ and integrate over $\Sigma$. We get the following inequality,
\ben
\int_{\Sigma}\frac{1}{N^{3}}+3\frac{|\nabla N|^{2}}{N^{4}}dv_{g}=\int_{\Sigma}\frac{|K|^{2}}{N^{2}}dv_{g}\leq (\int_{\Sigma}|K|^{6}dv_{g})^{\frac{1}{3}}(\int_{\Sigma}\frac{1}{N^{3}}dv_{g})^{\frac{2}{3}}.
\een

\n This shows that $1/N$ is controlled in $H^{1}_{g}$ and therefore in $L^{6}_{g}$ from above. Now compute the Laplacian of $1/N$
\be\label{D1/N}
\Delta \frac{1}{N}=\frac{2|\nabla N|^{2}}{N^{3}}+\frac{1}{N^{2}}-|K|^{2}\frac{1}{N}.
\ee

\n As $|K|^{2}$ is controlled in $L^{3}_{g}$ then from the Lapse equation we get that $\nabla N$ is controlled in $H^{1,3}_{g}$ and therefore controlled in $L^{q}_{g}$ for any $q>1$. As $1/N$ is controlled in $L^{6}_{g}$ we get that $2|\nabla N|^{2}/N^{3}$ is controlled in $L^{p}_{g}$ for any $p<2$. Therefore, standard elliptic estimates applied to the equation (\ref{D1/N}) (where the right hand side is thought as a non-homogeneous term) shows that $1/N$ is also controlled in $C^{0}$ by $diam_{g}(\Sigma),\ 1/Vol_{g}(\Sigma)$ and $Q_{0}$.\ep

\vs
This result can be easily adapted to the AF maximal setting but instead of working with $1/N$ one has to work with $\ln N$. There are some striking consequences of this Lemma that concern the space-time. The following corollary is easily deduced.

\begin{Corollary} (U.C) Let $M=[k_{0},k_{1}]\times \Sigma$. Suppose that $k_{0}<k_{1}<0$. Assume too that for every $k$ in $[k_{0},k_{1}]$ we have $Q_{0}(k)\leq \Lambda_{0}$, $diam_{g(k)}(\Sigma)\leq D_{0}$ and $Vol_{g(k)}(\Sigma)\geq V_{0}$. Then the reciprocal of the volume radius of $(M,{\bf g}=N^{2}dk^{2}+g(k))$ is controlled from above by an upper bound of $|k_{0}|,\ |1/k_{1}|,\ \Lambda_{0},\ 1/V_{0}$ and $D_{0}$. 
\end{Corollary}

\n In fact, as the $L^{2}_{\bf g}$-norm of the Riemann tensor associated to the metric ${\bf g}$ is controlled from above, we can add to the conclusions of the Corollary above that the $H^{2}$-harmonic radius of ${\bf g}$ is controlled from below too.

\subsubsection{\sf The space-time curvature and the Cauchy problem.}\label{SSS4.2.1}

A somehow folkloric conjecture in General Relativity asserts, roughly speaking, that the CMC Einstein flow keeps running until the first time at which the Bel-Robinson energy diverges. This continuity criteria is a difficult problem, but if correct, would prove to be rather useful in any intrinsic and geometric approach to the Einstein flow. At the moment this is an entirely open problem. The next Proposition, based on the conjectures in Section \ref{S2}, is a step forward in this important question. It shows essentially that the flow keeps running until the first time at which the $L^{4}$-norm of the space-time curvature with respect to the slices diverges. Note that the Bel-Robinson energy represents the $L^{2}$-norm of the space-time curvature with respect to the slices. To our knowledge this would be the first result of the kind in the field of Mathematical General Relativity. 

\begin{Proposition} Let $(g(k),K(k))$ be a smooth\footnote{This, or a similar condition (see the next footnote), is required.} ($C^{\infty}$) Einstein CMC flow over a compact three-manifold $\Sigma$ of arbitrary 
topology. Assume that the initial time is $k_{0}< 0$ and that the range of the flow is inside $(-\infty,0)$. Then the flow (in any direction: volume expansion or volume contraction) keeps running as long as $diam_{g(k)}(\Sigma)$, $\int_{\Sigma}|E|^{4}+|B|^{4}dv_{g}$ and $|k|+1/|k|$ remain bounded. 
\end{Proposition}

\n {\it Sketch of the proof.} 

First note that because the {\it reduced volume} ${\mathcal{V}}=-k^{3}Vol_{g}(\Sigma)$ is monotonically decreasing in the expanding direction (i.e. as $k$ increases) the volume in the expanding direction is obviously bounded below and
is bounded above by $V(k)\leq (\frac{k_{0}}{k})^{3}V(k_{0})$ while in the contracting direction is obviously bounded above and bounded below by $V(k)\geq (\frac{k_{0}}{k})^{3}V(k_{0})$. Therefore from the bound (\ref{EE1}) and the Conjecture \ref{C1} we conclude that $1/\nu_{g(k)}(\Sigma)$ remains controlled from above as long as $diam_{g(k)}(\Sigma)$,
$\int_{\Sigma}|E|^{4}+|B|^{4}dv_{g}$ and $|k|+1/|k|$ are bounded from above (and naturally control the reciprocal of the $H^{2}$-harmonic radius from above). The idea now is to show that they also control the first and second order Bel-Robinson energy. As was shown in \cite{ReirisCMC} this is enough to control the evolution of the higher order Bel-Robinson energies. We deduce then from Lemma \ref{Intr1} that the flow is controlled in $C^{\infty}$ (with respect to suitable harmonic coordinates)\footnote{It is not necessary to appeal to this more elaborate result, indeed imposing only that the flow is in $C^{1}(H^{3}\times H^{2}_{g})$, it is not difficult to show using standard formulations of the Cauchy problem in $H^{3}\times H^{2}$, that the flow remains controlled (in this norm) as long as $1/\nu(\Sigma)+Q_{0}+Q_{1}+diam_{g_{k}}(\Sigma)$ remains bounded above.} . Naturally $Q_{0}$ is automatically controlled. To analyze $Q_{1}$ we look at the Gauss equation (\ref{Gausseq}). We get
\be\label{EQ1}
\frac{\partial}{\partial t}Q_{1}\leq c_{1}\|N\|_{C^{0}}\|{\bf \Pi}\|_{C^{0}}Q_{1}+
c_{2}\|N\|_{C^{0}}Q_{1}^{\frac{1}{2}}\|{\bf J}({\bf W}_{1})\|_{L^{2}_{g}}.
\ee

\n From the elliptic system (\ref{EK1})-(\ref{EK2}) and the bound (\ref{EE1}) we get that $\int_{\Sigma}|E|^{4}+|B|^{4}dv_{g}$, $diam_{g}(\Sigma)$ and $|k|+1/|k|$ (below called the main variables) control $\|K\|_{H^{1,4}_{g}}$ and therefore from Sobolev embeddings we get control on $\|K\|_{C^{0}_{g}}$. From the lapse equation (\ref{LE}) we get control on $\|N\|_{C^{1}_{g}}$ and therefore on $\|{\bf \Pi}\|_{C^{0}_{g}}$. From standard elliptic estimates applied to the elliptic system (\ref{eq1})-(\ref{eq2}) we get that $E({\bf W}_{1})$ and $B({\bf W}_{1})$ 
satisfy the bound
\be\label{ESE1}
\|E\|_{H^{1}_{g}}+\|B\|_{H^{1}_{g}}\leq c(Q_{0}^{\frac{1}{2}}+Q_{1}^{\frac{1}{2}}),
\ee

\n where $c$ depends on the main variables. A computation gives
\ben
{\bf J}({\bf W}_{1})={\bf \Pi}^{ab}*\bn_{b}{\bf W}_{acde}+T*{\bf Rm}*{\bf Rm},
\een

\n and we can write (denote by $P_{a}^{\ a'}$ the projection operator into the slices of the CMC foliation)
\be\label{ESE2}
P_{a}^{a'}\nabla_{a'}{\bf W}_{0}=K*\epsilon*\epsilon*E+\epsilon*\epsilon*\nabla E+K*\epsilon*B*T+\epsilon*\nabla B*T
+\epsilon*B*K+\nabla E*T*T+E*T*K.
\ee

\n where $*$ represents a tensor contraction or operation. Combining the estimate (\ref{ESE1}) and the expression (\ref{ESE2}) inside equation (\ref{EQ1}) gives 
\ben
\frac{d}{dt}Q_{1}\leq c(Q_{1}+Q_{1}^{\frac{1}{2}}).
\een

\n where $c$ depends on the main variables. This equation implies that $Q_{0}$ and $Q_{1}$ remain bounded as along as the main variables remain bounded, thus proving the Proposition.\ep

\subsection{\sf (Potential) applications to the formation of singularities: mean-curvature and velocity dominated singularities}\label{SS4.3}

\subsubsection{\sf Blow-up limits at points of maximal curvature.}\label{SSS4.3.1}

In this section we will explain some ideas showing how to use the conjectures that were raised in Section \ref{S2} to study the blow up cosmological CMC solutions. Blow ups are required to investigate for instance cosmological singularities, singularities inside black holes, or simply but more generally, to investigate the structure of the Einstein flow. In the first two situations at least, the quantities that diverge are (possibly) the Bel-Robinson energy $Q_{0}$ and (certainly) the mean curvature $k$. In the third scenario one may consider situations where $k$ does not diverge but $Q_{0}$ does. Thus to blow up a solution one has to pay attention to the concentration of the Bel-Robinson energy, that is, to the Bel-Robinson radius $r_{Q_{0}}$, and the mean curvature $k$, simultaneously. Note however that in order to use Conjecture \ref{C3} it is required a priori knowledge on the curvature radius $r_{C}$. As it is stated, such Conjecture is not asserted to be valid if we replace $r_{C}$ by $r_{Q_{0}}$ (we believe that this is a necessary and interesting question). For this reason we will work below with the curvature radius $r_{C}$ and the {\it magnetic radius} $r_{B}$. 

Conjecture \ref{C3} allows us to take blow up limits at points of maximal curvature (or minimal curvature radius). Without deeper understanding of the dynamics of the Einstein flow, its scope is, at the moment, restricted only to this case. To extract, in the blow up limit, a complete {\it space-time solution}, the factor by which we have to blow up states is not necessarily $1/r_{C}^{2}$ as we need $r_{C}(p)$, $r_{B}(p)$ and $k$ to be finite in the limit (where $p$ is the limit of the blown up points). For the sake of concreteness we will illustrate this point and further considerations on the blow-up technique, by considering two situations that we will call {\it mean-curvature dominated} and {\it velocity dominated}. 

Let $(p_{i},k_{i})$, with $k_{i}$ monotonically increasing (decreasing), be a sequence of space-time points such that $r_{C}(p_{i},k_{i})$ {\it is minimal}, i.e. $r_{C}(p,k)\geq r_{C}(p_{i},k_{i})$ for any other space-time point $(p,k)$ with $k\leq (\geq) k_{i}$. The flow is said to be {\it mean-curvature dominated\footnote{This definition is ours.} at $(p_{i},k_{i})$ if $\lim r_{C}(p_{i},k_{i})|k_{i}|\geq \Lambda_{1}$ and $\lim r_{B}(q,k_{i})/r_{C}(q,k_{i})\geq \Lambda_{2}>0$, for all $q$ in $\Sigma$, for some fixed $\Lambda_{1}>0$ and $\Lambda_{2}>0$. It is said to be velocity dominated if $\lim r_{C}(p_{i},k_{i})|k_{i}|\rightarrow \infty$ and $\lim r_{B}(q,k_{i})/r_{C}(q,k_{i})\geq \Lambda_{2}>0$, for all $q$ in $\Sigma$, for some fixed $\Lambda_{2}>0$}.

In any of these cases one has to blow up the (pointed) state $((g,K)(k_{i}),p_{i})$ by the factor $1/k^{2}_{i}$. To apply Conjecture \ref{C3} and get a complete {\it limit state} however, it is necessary that the volume radius of the blown up solution at the point $(p_{i},k_{i})$ is controlled from below and away from zero. Ideally there should exist a weak quantity, perhaps not involving the curvature, that would control the volume radius from the data on an initial slice. In some circumstances the space-time volume, when used adequately, provides an interesting alternative (see the comments at the end of the next section). Let us show now that, granted the validity of Conjecture \ref{C3}, the blow up limit of a mean-curvature dominated is a well defined complete state and that the limit of a velocity dominated sequence of states is a {\it Kasner state}. Recall that a $(\alpha_{1},\alpha_{2},\alpha_{3})$-{\it Kasner solution} is given explicitly by 
\ben
{\bf g}=-dt^{2}+a_{1}^{2}t^{2\alpha_{1}}d\theta_{1}^{2}+a_{2}^{2}t^{2\alpha_{2}}d\theta_{2}^{2}+a_{3}^{2}t^{2\alpha_{3}}d\theta_{3}^{2},
\een

\n where $(\alpha_{1},\alpha_{2},\alpha_{3})$ satisfy $\alpha_{1}+\alpha_{2}+\alpha_{3}=1$ and $\alpha_{1}^{2}+\alpha_{2}^{2}+\alpha_{3}^{2}=1$. A $(\alpha_{1},\alpha_{2},\alpha_{3})$-{\it Kasner state} is simply a state (at a given $t$) of a Kasner $(\alpha_{1},\alpha_{2},\alpha_{3})$-solution (regardless of the topology). 

\begin{Proposition} (U.C) Let $\{(p_{i},k_{i})\}$ be a mean curvature dominated sequence of space-time points. Consider the blown up sequence of states $(\tilde{g}_{i},\tilde{K}_{i})=(\frac{1}{k_{i}^{2}}g(k_{i}),\frac{1}{k_{i}}K(k_{i}))$. Suppose that $\nu_{\tilde{g}_{i}}(p_{i})\geq \nu_{0}>0$. Then there is a subsequence converging to a CMC state $(\tilde{g}_{\infty},\tilde{K}_{\infty})$ with $k_{\infty}=1$. If in addition the sequence is velocity dominated then the limit state is a Kasner state.
\end{Proposition}

\n {\it Sketch of the proof:}

	First note that if $r_{C}(p_{i},k_{i})|k_{i}|\geq \Lambda_{1}>0$ then $\tilde{r}_{C,i}(p_{i})\geq \Lambda_{1}$. From this and the definition of mean-curvature dominated ($p_{i}$ is of minimal curvature radius) we deduce that $\tilde{r}_{C,i}\geq \Lambda_{1}$. Thus from Conjecture \ref{C3} one knows that there is a subsequence of $\{(\tilde{g}_{i},p_{i})\}$ converging to a complete limit $(\tilde{g}_{\infty},p_{\infty})$. Now we use the second condition in the definition of mean-curvature dominated to conclude that $\tilde{r}_{B,i}(q)\geq \Lambda_{1}\Lambda_{2}$ and therefore $\tilde{r}_{B,\infty}\geq \Lambda_{1}\Lambda_{2}$. Given $d_{0}>0$ apply Lemma \ref{LF} to any sequence $\{q_{i}\}$ of points such that $dist_{\tilde{g}_{i}}(q_{i},p_{i})\leq d_{0}$ to conclude that the radius $\tilde{r}_{K,i}(q)$ is controlled from below and away from zero for points $q$ at distance from $p$ less or equal than $d_{0}$. As $d_{0}$ is arbitrary we conclude that $\tilde{r}_{K,\infty}(q)$ is positive for any point $q$ in the limit manifold. Elliptic regularity applied to the elliptic system (\ref{EK1})-(\ref{EK2}) then shows that $\tilde{K}_{\infty}$ is a well defined state in $H^{1}$. 
	
	Now let us suppose that the sequence of points is in addition velocity dominated. This implies in particular that the limit state will has $\tilde{r}_{C,\infty}=\infty$, thus showing that it is flat. From the definition of velocity dominated one knows too that $\tilde{r}_{B,\infty}=\infty$ and therefore $B_{\infty}=0$. From the energy constraint we know that $|\tilde{K}_{\infty}|^{2}=1$. On the other hand using equation (\ref{Simons2}) when the manifold is flat and the magnetic field is zero gives
\ben
\Delta \tilde{K}_{\infty} =0,
\een

\n and therefore
\ben
\Delta R_{\infty}=\Delta |\tilde{K}_{\infty}|^{2}=2<\nabla \tilde{K}_{\infty},\nabla \tilde{K}_{\infty}>=0.
\een

\n Thus $\tilde{K}_{\infty}$ is parallel and therefore constant. This is enough to conclude that the state is Kasner.\ep 

Observe that we have not claimed that the space-time metric converges (weakly) into a solution of the Einstein equations. While this should be the case (because the sequence of blow up points is of maximal curvature), a detailed discussion of the assertion would require an independent treatment that we will not pursue here.  
 	
\subsubsection{\sf Example: the interior Schwarzschild.}\label{SSS4.3.2}

In this section we provide a standard example of a solution that is mean-curvature dominated and more in particular velocity dominated. The solution is the interior of the Schwarzschild and indeed, because of its symmetries, every sequence of points approaching the singularities (there are two) is of maximal curvature (minimal curvature radius). Because the solution is simple and exact, we will avoid detailed computations (which can be given at any extent) and use elementary approximations to justify the claims. The {\it interior Schwarzschild} is the solution  
\ben
{\bf g}=(\frac{2M}{t}-1)dr^{2}-\frac{1}{(\frac{2M}{t}-1)}dt^{2}+t^{2}d\Omega^{2},
\een

\n on $r\in [r_{0},r_{1})\times S^{2}$ and $r_{0}$ is identified with $r_{1}$. Now we want to study the singularities at $t=0$ and $t=2M$. A computation 
gives
\ben
k=-\frac{\sqrt{2M-t}}{t^{\frac{3}{2}}}(\frac{M}{2M-t}-2)).
\een

\n When $t\sim 0$, $k\sim \frac{3}{2}\sqrt{2M}\frac{1}{t^{\frac{3}{2}}}$ and when $t\sim 2M$, $k\sim -M^{-\frac{1}{2}}/(\sqrt{2M-t})(2)^{\frac{3}{2}}$.

\n {\it Case I: t$\rightarrow$ 0}. First note that in this case $r_{C}\sim t$ but $k\sim 1/t^{\frac{3}{2}}$ so $r_{C}.k\rightarrow \infty$ as $t\rightarrow 0$. When $t\rightarrow 0$ we have
\ben
{\bf g}\sim \frac{2M}{t}dr^{2}-\frac{t}{2M}dt^{2}+t^{2}d\Omega^{2}.
\een

\n Make $t'=t^{\frac{3}{2}}$. We can write the metric in the form 
\ben
{\bf g}\sim-dt'^{2}+t'^{-\frac{2}{3}}dr'^{2}+t'^{\frac{4}{3}}d\Omega'^{2}.
\een

\n where $r'$ and $\Omega'$ are numerically scaled quantities, which after scaling by $1/k_{0}^{2}$ ($k_{0}$ is the mean curvature at the time $t_{0}$) and redefining time by $t''=t'/t'_{0}$ and also redefining coordinates (by just a factor) we get
\ben
{\bf g}\sim-dt''+t''^{-\frac{2}{3}}d\theta_{1}^{2}+ct''^{\frac{4}{3}}(d\theta_{2}^{2}+d\theta_{3}^{2}).
\een

\n which is a Kasner $(-1/3,2/3,2/3)$.  

\vspace{0.2cm}
\n {\it Case II: t$\rightarrow$ 2M}.

\n We have
\ben
{\bf g}=\frac{2M-t}{t}dr^{2}+\frac{t}{(t-2M)}dt^{2}+t^{2}d\Omega^{2}.
\een

\n and making $t'=2M-t$
\ben
{\bf g}=\frac{t'}{(2M-t')}dr'^{2}+(2M-t')^{2}d\Omega^{2}-\frac{(2M-t')}{t'}dt'^{2}.
\een

\n We have $k_{0}\sim -1/t_{0}'^{\frac{1}{2}}$. Note that in this case we have $r_{C}\sim t'^{\frac{1}{2}}$. Blowing up by $1/k_{0}^{2}$,
redefining time by $t''=(t'/t'_{0})^{\frac{1}{2}}$ and also coordinates by a factor, we get
\ben
{\bf g}\sim t''^{2}d\theta_{1}^{2}+d\theta_{2}^{2}+d\theta_{3}^{2}-dt''^{2}.
\een

\n The solution thus is getting closer to the Kasner $(1,0,0)$ (as it has to be because it is a Cauchy horizon).
 
It is not a coincidence that in this particular example the scale invariant combination $\nu_{g(k)}(k).|k|$ is uniformly bounded below and away from zero (it actually diverges) when the mean curvature $k$ diverges, i.e. when $t\rightarrow 0$ or $t\rightarrow 2M$. Indeed this is a consequence of the particular behavior of the slices with respect to a given initial slice, say the one with $k=0$. In rough terms the basic requirement is that the    
slices at the sequence of blow up points do not ``twist" with respect to the initial slice. A precise definition of ``twist" can indeed be given. One possible way is as follows. The CMC slices of a sequence of blow up points $(p_{i},k_{i})$  do not twist near the blow up point iff there are $V_{0}>0$ and $\alpha_{0}>0$ and, for every $i$, a set of length maximizing time-like geodesics $\{\gamma\}_{i}$ joining $(p_{i},k_{i})$ to the initial slice such that 
\begin{enumerate}
\item for each $i$, the space-time volume of the space-time set formed by all the geodesics is greater or equal than $V_{0}$ (independent on $i$),
\item for each $i$, $|<\gamma'(p_{i},k_{i}),T(p_{i},k_{i})>|\leq \alpha_{0}$, where $T$ is the normal vector to the foliation.
\end{enumerate}

\n It can be shown through a standard use of the Bishop-Gromov volume comparison (in the Lorentzian setting) that if a sequence of points is mean curvature dominated and enjoying the ``no twist" property then $\nu_{g_{i}}(p_{i},k_{i})|k_{i}|\geq \nu_{0}>0$ where $\nu_{0}$ is fixed but depends on the solution. This ``no twist" property is not difficult to guess in the particular example of the interior Schwarzschild example where the CMC slices are equidistant.

\end{document}